\documentstyle[sprocl,epsfig,epsf,wrapfig,rotating]{article}

\def\be{\begin{equation}}
\def\ee{\end{equation}}
\def\bea{\begin{eqnarray}}
\def\eea{\end{eqnarray}}

\def\sqee{\sqrt{s}_{\rm ee}}

\def\gg{\gamma\gamma}

\def\ee{\mbox{e}^+\mbox{e}^-}

\def\ETJET{E^{\rm jet}_T}

\def\qqbar{\mbox{q}\overline{\mbox{q}}}

\def\xgp{x_{\gamma}^+}
\def\xgm{x_{\gamma}^-}
\def\xgpm{x_{\gamma}^{\pm}}
\def\etajet{\eta^{\rm jet}}

\def\Zzero{\ifmmode {{\mathrm Z}^0} \else {${\mathrm Z}^0$} \fi}
\def\ppbar{\overline{\mbox p}\mbox{p}}

\def\cost{\cos\theta^*}
\def\sigmagg{\sigma_{\gg}}
\def\pt{p_{\rm T}}
\def\dspt{{\rm d}\sigma/{\rm d}p_{\rm T}} 

\begin{document}

\setcounter{footnote}{0}
\renewcommand{\thefootnote}{\fnsymbol{footnote}}
\title{TESTING QCD IN PHOTON-PHOTON INTERACTIONS\footnotemark[1] }

\author{STEFAN S\"OLDNER-REMBOLD}

\address{for the OPAL collaboration\\ Albert-Ludwigs-Universit\"at Freiburg\\ 
Hermann-Herder-Str.3,\\ D-79104 Freiburg, Germany\\E-mail: 
soldner@ruhpb.physik.uni-freiburg.de} 

\vspace{-0.3cm}

\maketitle\abstracts{
At high energies photon-photon interactions are 
dominated by quantum fluctuations
of the photons into fermion-antifermion pairs and into
vector mesons. This is called photon structure. 
Electron-positron collisions at LEP are an ideal laboratory
for studying photon structure and for testing QCD.} 
\setcounter{footnote}{0}
\renewcommand{\thefootnote}{\fnsymbol{footnote}}
\footnotetext[1]{Topical Lecture given at the Lake Louise Winter Institute, 
Canada, February 15-21, 1998}

\setcounter{footnote}{0}
\renewcommand{\thefootnote}{\alph{footnote}}
\section{Electron-photon scattering}
\begin{wrapfigure}[15]{l}{2.5in}
\epsfig{file=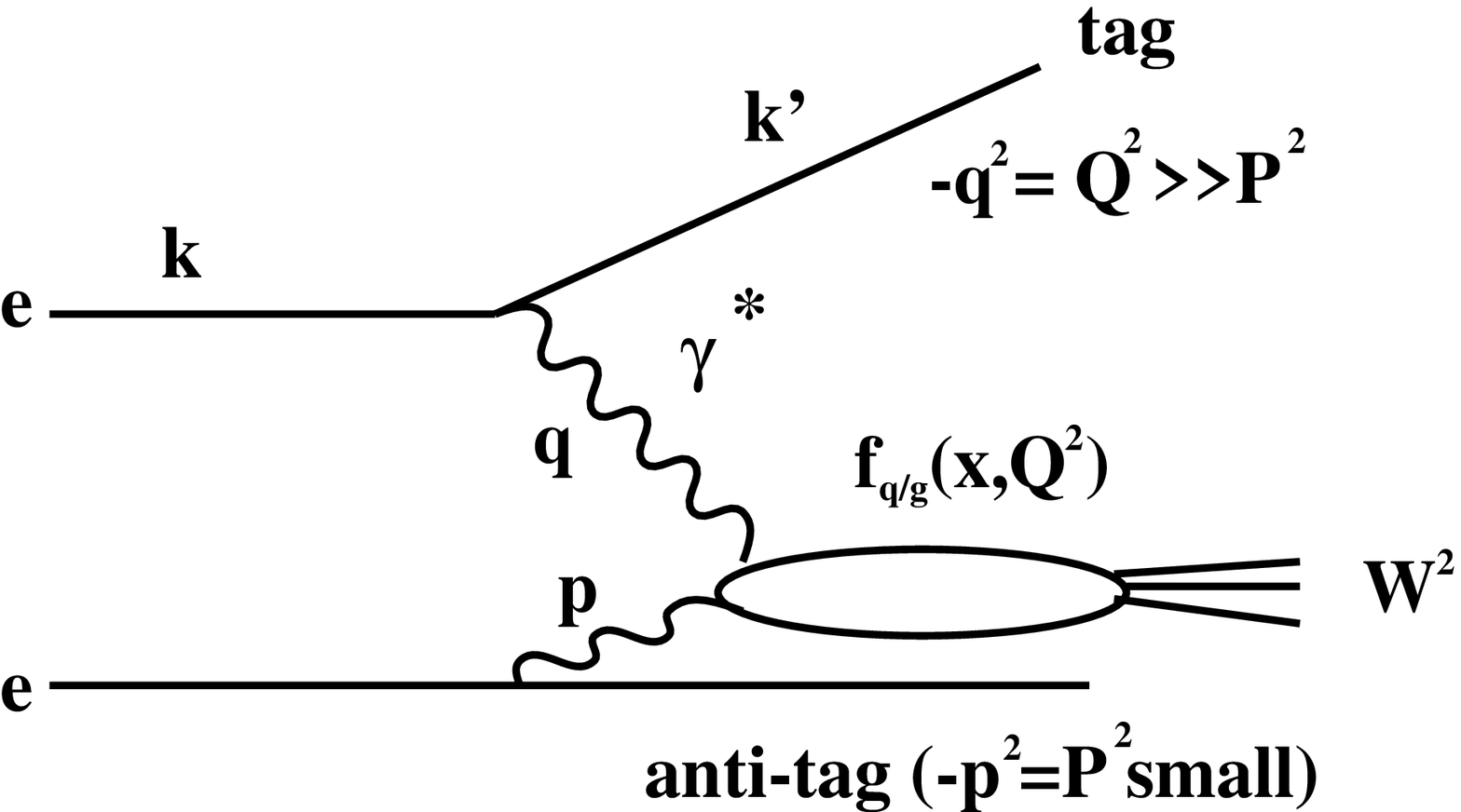,width=2.5in} 
\caption{\label{fig-egfig}
Deep-inelastic e$\gamma$ scattering:
$k(k')$ denotes the four-momentum of the incoming (scattered) electron
and $q(p)$ the four-momentum of the virtual (quasi-real) photon.}
\end{wrapfigure}

If one of the scattered electrons in $\ee$ collisions is detected (tagged), 
the process $\ee \rightarrow \ee + \mbox{hadrons}$ (Fig.~\ref{fig-egfig}) 
can be regarded as deep-inelastic scattering of an 
electron\footnote{In this paper positrons 
are also referred to as electrons}~on 
a quasi-real photon which has been radiated by the other electron
beam. The cross-section is written as
 \begin{equation}
  \frac{{\rm d}^2\sigma_{\rm e\gamma\rightarrow {\rm e+hadrons}}}{{\rm d}
x{\rm d}Q^2}
 =\frac{2\pi\alpha^2}{x\,Q^{4}}
  \left[ \left( 1+(1-y)^2\right) F_2^{\gamma}(x,Q^2) - y^{2}
F_{\rm L}^{\gamma}(x,Q^2)\right],
\label{eq-eq1}
 \end{equation}
where $\alpha$ is the fine structure constant and 
$$Q^2=-q^2=-(k-k')^2$$ 
is the negative four-momentum squared of the virtual photon $\gamma^*$ and 
$$x=\frac{Q^2}{2p\cdot q}=\frac{Q^2}{Q^2+W^2+P^2} \;\;\;\mbox{;}\;\;\;
y=\frac{p\cdot q}{p\cdot k}$$
are the usual dimensionless variables of deep-inelastic scattering.
$W^2=(q+p)^2$ is the squared invariant mass of the hadronic final state.
The negative four-momentum squared, $P^2=-p^2$,
of the quasi-real target photon is approximately zero.
In leading order (LO) the photon structure function $F_2^{\gamma}(x,Q^2)$ 
is related to the sum over the quark densities of the 
photon
weighted by the quark charge $e_{\rm q}$
$$F_2^{\gamma}(x,Q^2)=2x\sum_{\rm q} e^2_{\rm q} 
f_{\rm q/\gamma}(x,Q^2)$$ 
with $f_{\rm q/\gamma}(x,Q^2)$ being the probability to
find a quark flavour q with the momentum fraction $x$ 
(sometimes denoted by $x_{\gamma}$) in the photon. 
For measuring $F_2^{\gamma}(x,Q^2)$ the values 
of $Q^2$ and $y$ can be reconstructed from the energy,
$E_{\rm tag}$,
and the angle, $\theta_{\rm tag}$, of the tagged electron and the beam
energy $E_{\rm beam}$:
$$Q^2\approx 2E_{\rm beam} E_{\rm tag}(1-\cos\theta_{\rm tag})$$
$$y\approx 1-\frac{E_{\rm tag}}{E_{\rm beam}}\cos^2\frac{\theta_{\rm tag}}{2}.$$
In order to identify an electron in the detector, 
$E_{\rm tag}$ has to be large, i.e.~$y^2\ll 1$. 
The contribution of the term proportional to the 
longitudinal structure function $F_{\rm L}^{\gamma}$ is therefore negligible
(Eq.~\ref{eq-eq1}).
 
The reconstruction of $x$, however, relies heavily on the 
measurement of the invariant mass $W$ from the energies $E_{\rm h}$ and
momenta $\vec{p}_{\rm h}$ of the final state hadrons h:
$$W^2=\left( \sum_{\rm h} E_{\rm h} \right)^2-\left( \sum_{\rm h}
\vec{p}_{\rm h}\right )^2.$$
Unfolding of the $x$ dependence of $F_2^{\gamma}$
requires that the hadronic final state
is well measured and well simulated by the Monte Carlo (MC) models.

\subsection{The photon structure function $F_2^{\gamma}$}
\label{sec-egamma}
Even though the concept of the photon structure function 
$F_2^{\gamma}$ has
been developed in analogy to the formalism of the nucleon structure
functions $F_2^{\rm N}$, there are important differences: 
$F_2^{\gamma}(x,Q^2)$ increases with $Q^2$ for all $x$ and this
positive scaling violation is expected already within the parton model. 
Furthermore, $F_2^{\gamma}$ is large for high $x$, whereas
$F_2^{\rm N}$ decreases at large $x$. These differences are due
to the additional perturbative $\gamma\rightarrow\qqbar$ splitting
which does not exist for the nucleon. 

For large $x$ and asymptotically large $Q^2$ the value of
$F_2^{\gamma}$ can therefore be calculated from perturbative 
QCD~\cite{bib-witten}. The next-to-leading order (NLO) 
result~\cite{bib-buras} can be written as
\begin{equation}
\frac{F_2^{\gamma}}{\alpha}=\frac{a(x)}{\alpha_{\rm s}(Q^2)}+b(x),
\end{equation}
where $a(x)$ and $b(x)$ are calculable functions which diverge
for $x\rightarrow 0$ and $\alpha_{\rm s}$ is the strong coupling constant.
The first term corresponds to the LO
result by Witten~\cite{bib-witten}. The measurement
of $F_2^{\gamma}$ could be a direct measurement of $\Lambda_{\rm QCD}$
if it were not for the large non-perturbative contributions
due to hadronic states. 

\unitlength1cm
\begin{figure}[htbp]
\begin{picture}(15.0,6.2)
\put(0,3.25){\begin{tabular}{cc}
\epsfig{file=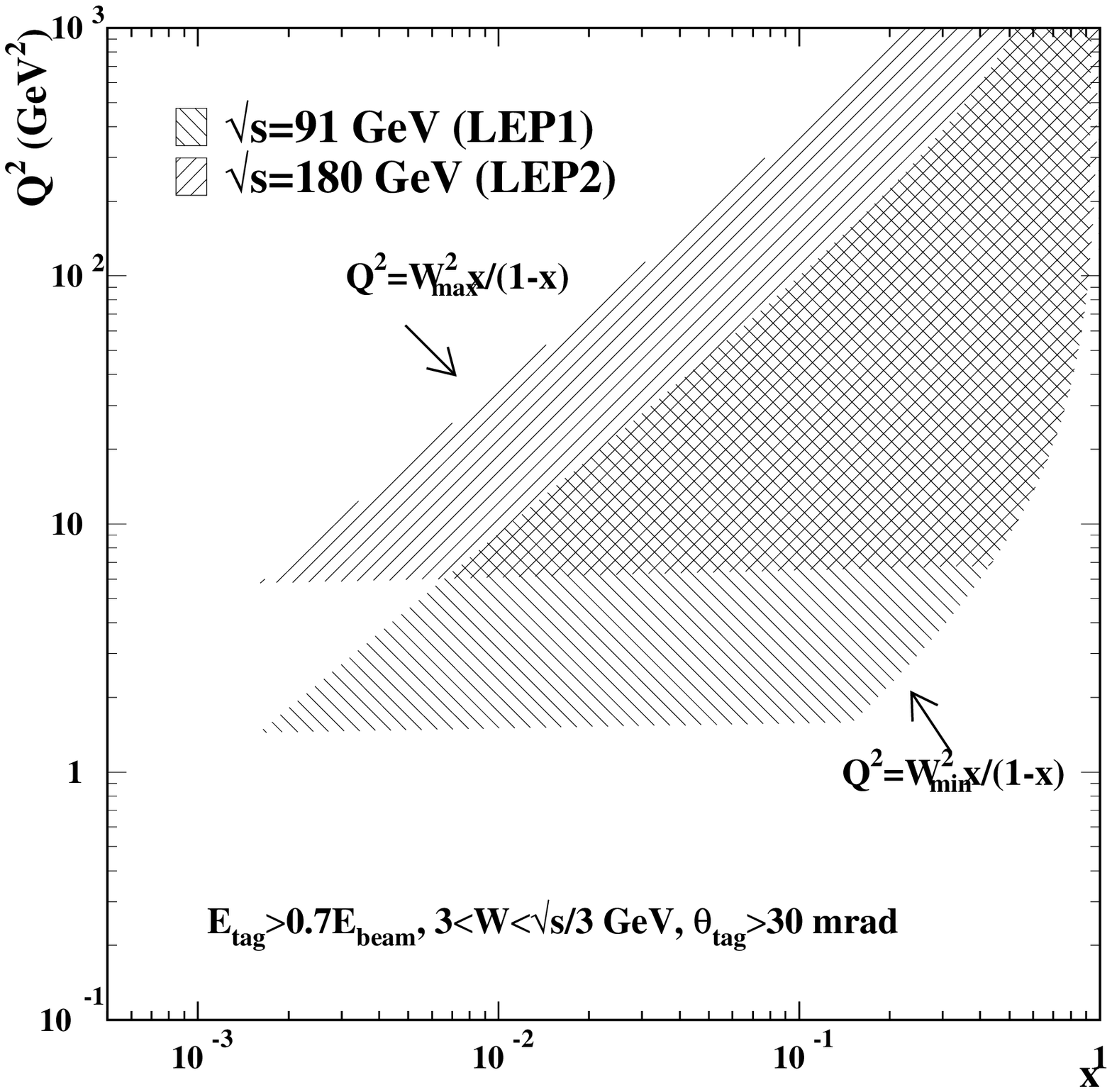,width=0.46\textwidth,height=6.2cm}
 &
\epsfig{file=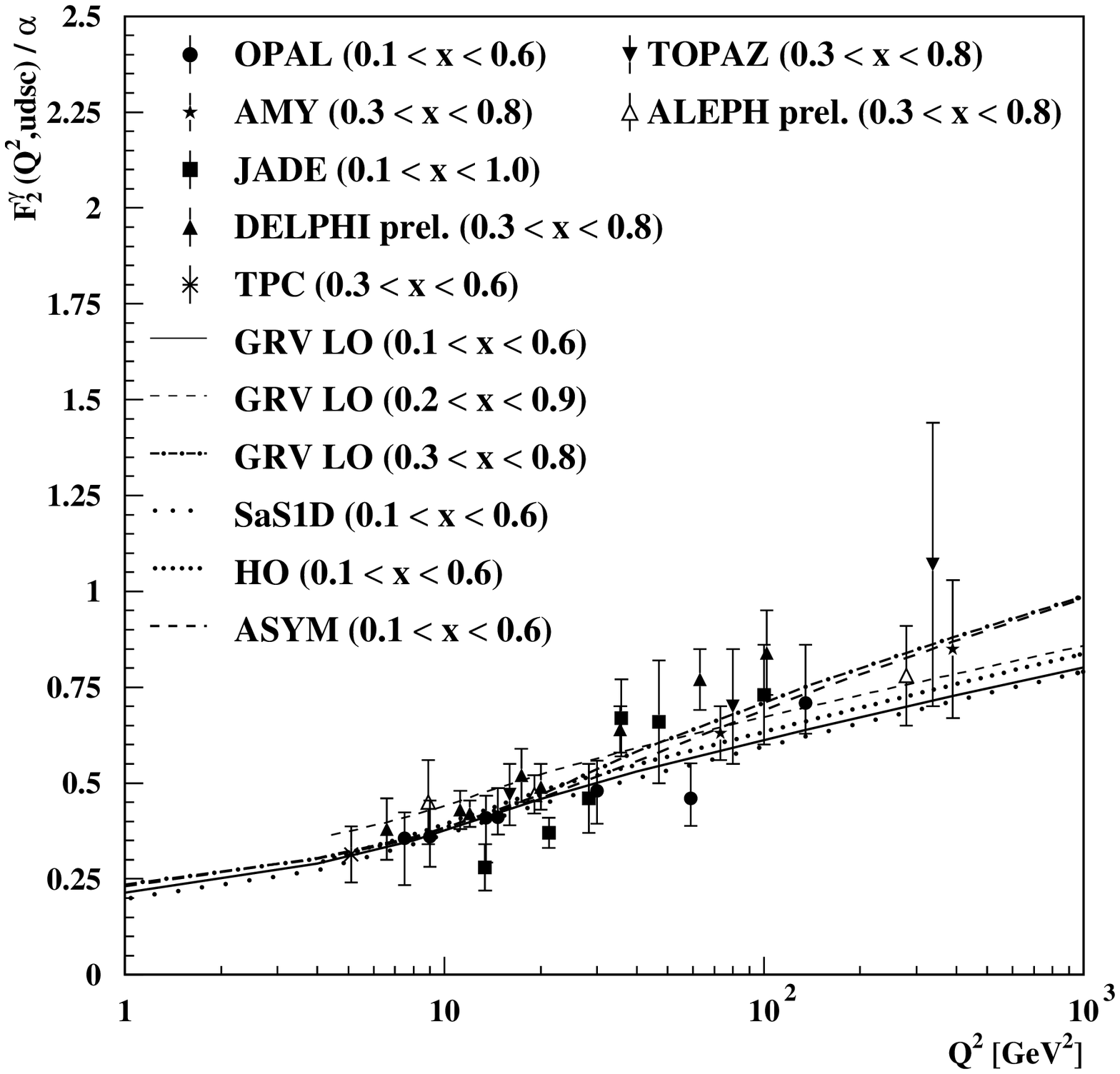,width=0.46\textwidth,height=6.2cm} 
\end{tabular}}
\put(5.0,1.2){(a)}
\put(11.1,1.2){(b)}
\end{picture}
\caption{\label{fig-cov}
a) Kinematical coverage of the ($Q^2,x$) plane at LEP1 and LEP2. 
b) The photon structure function $F_2^{\gamma}/\alpha$
as a function of $Q^2$.} 
\end{figure}

$F_2^{\gamma}(x,Q^2)$ can be measured at LEP
in the range $x>10^{-3}$ and $1<Q^2<10^3$~GeV$^2$ (Fig.~\ref{fig-cov}a). 
The $Q^2$ evolution of $F_2^{\gamma}$ is
shown in Fig.~\ref{fig-cov}b using the currently available
$F_2^{\gamma}$ measurements for 4 active flavours. The data are compared to
the LO GRV~\cite{bib-grv} and the SaS-1D~\cite{bib-sas}
parametrisations, and to a higher
order (HO) prediction based on the NLO GRV parametrisation for
light quarks and on the NLO charm contribution calculated in
Ref.~\cite{bib-laenen}. The data are measured in different $x$ ranges. 
The comparison of the LO GRV curves
for these $x$ ranges shows that for $Q^2>100$~GeV$^2$ significant
differences are expected.
An augmented asymptotic prediction for $F_2^{\gamma}$ is also
shown. The contribution to $F_2^{\gamma}$ from the three light flavours is 
approximated by Witten's LO asymptotic form~\cite{bib-witten}.
This has been augmented by adding a charm contribution 
from the Bethe-Heitler formula~\cite{WIT-7601}, and an estimate of 
the hadronic part of $F_2^{\gamma}$ based
on the hadronic part of the LO GRV parametrisation.
In the region of medium $x$ values studied here, this asymptotic prediction 
in general lies higher than the GRV and SaS predictions but it is still 
in agreement with the data.
The importance of the hadronic contribution to $F_2^{\gamma}$
decreases with increasing $x$ and $Q^2$, 
and it accounts for only 15~\% of $F_2^{\gamma}$ at 
$Q^2= 59$~GeV$^2$ and $x = 0.5$.
As predicted by QCD the evolution of $F_2^{\gamma}$ leads
to a logarithmic rise with $Q^2$, but theoretical and experimental
uncertainties are currently too large for a precision test of perturbative QCD.
\begin{figure}[htbp]
   \begin{center}
      \mbox{
 \epsfig{file=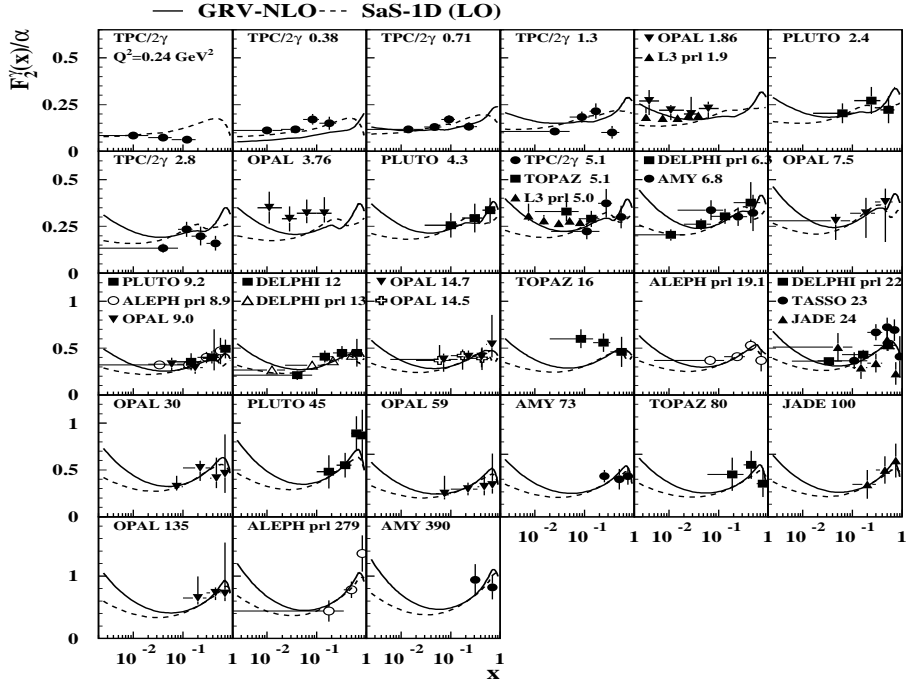,width=1.0\textwidth,height=9cm}
           }
   \end{center}
\caption{\label{fig-sf}
Measurements of the photon structure function $F_2^{\gamma}$
in bins of $x$ and $Q^2$. }
\end{figure}

All currently available $F_2^{\gamma}$ measurements~\cite{bib-f2g}
are compared to the NLO GRV~\cite{bib-grv} 
and the LO SaS-1D~\cite{bib-sas} parametrisation in Fig.~\ref{fig-sf}. 
If the photon is purely hadron-like at low $x$,
a rise of the photon structure function
is expected at low $x$ for not too small 
$Q^2$, similar to the rise of the proton structure function.
Only with the complete LEP2 data will it
be possible to access regions in $x$ and $Q^2$ where the
rise of $F_2^{\gamma}$ could really be observed.
An interesting low $x$ measurement of $F_2^{\gamma}$ by OPAL lies
in the ranges $2.5\times10^{-3}<x<0.2$ and $1.1<Q^2<6.6$~GeV$^2$.
L3 has recently presented their
first $F_2^{\gamma}$ measurement for $Q^2=1.9$ and 
$5.0$~GeV$^2$. These measurements are
consistent with a possible rise within large errors. 

\newpage
\section{Jet production and NLO calculations }
\label{sec-jet}
\begin{wrapfigure}[20]{r}{0.47\textwidth}
\epsfig{file=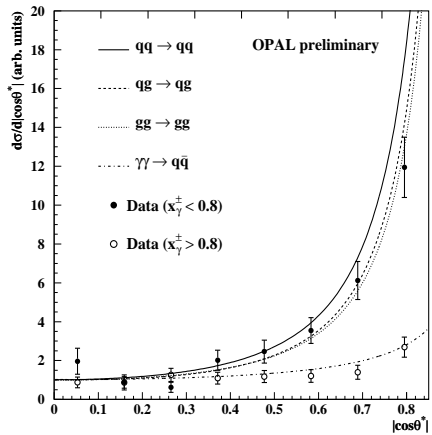,width=0.47\textwidth,height=5.7cm}
\caption{\label{fig-theta} 
Angular distribution for $\xgpm>0.8$ and $\xgpm<0.8$.
The data are compared to LO QCD calculations. 
The data are normalised
to 1 in the first three bins and the curves are normalised to 
$\cos(0)=1$. 
}
\end{wrapfigure}
If the virtualities $Q^2$ and $P^2$ 
are approximately zero, i.e.~both photons are quasi-real,
LEP2 is a $\gg$ collider with $\gg$ centre-of-mass energies
in the approximate range $10<W<120$~GeV. 

In LO different event classes can be defined
in $\gg$ interactions. The photons can either
interact as bare photons (``direct'') or as hadronic fluctuation
(``resolved''). 
Direct and resolved events can be separated by measuring
the fraction $\xgpm$ of the photon's momentum participating in the 
hard interaction for the two photons.
Ideally, the direct events are expected to have $\xgpm=1$,
whereas for double-resolved events both values $\xgp$ and $\xgm$
are expected to be much smaller.

For a given parton centre-of-mass energy the cross-sections
vary only with the parton scattering angle $\theta^{*}$.
The leading order direct process $\gg \rightarrow \qqbar$ is mediated
by $t$-channel spin-$\frac{1}{2}$ 
quark exchange which leads to an angular dependence
$\propto (1-\left|\cos\theta^{*2}\right|)^{-1}$.
In double-resolved processes all matrix elements
involving quarks and gluons have to be taken into account,
with a large contribution from spin-$0$ gluon exchange. 
After adding up all relevant processes, perturbative QCD
predicts an angular dependence of approximately $\propto
(1-\left|\cost\right|)^{-2}$.
Figure~\ref{fig-theta} shows the corrected $|\cost|$ distribution
of dijet events with $\xgpm>0.8$ and with $\xgpm < 0.8$
compared to the calculation for various LO matrix elements which
qualitatively reproduce the data.

NLO jet cross-sections for $\gg$
interactions have been calculated by many 
authors~\cite{bib-kleinwort,bib-aurenche}
using the cone jet finding 
algorithm~\cite{bib-cone}.
The transverse momentum $p_{\rm T}$
of the final-state partons (or the jet) defines the hard scale. 
The jet cross-section is written as a convolution of 
the parton density of the photon 
with the matrix elements for the
scattering of two partons.
In the kinematic range covered by LEP
the $F_2^{\gamma}$ measurements are mainly probing
the quark content of the photon, whereas
jet production can be used
to constrain the relatively unknown gluon distribution in the photon.

\begin{figure}[htbp]
\begin{picture}(15.0,6.0)
\put(0,3.0){\begin{tabular}{cc}
\epsfig{file=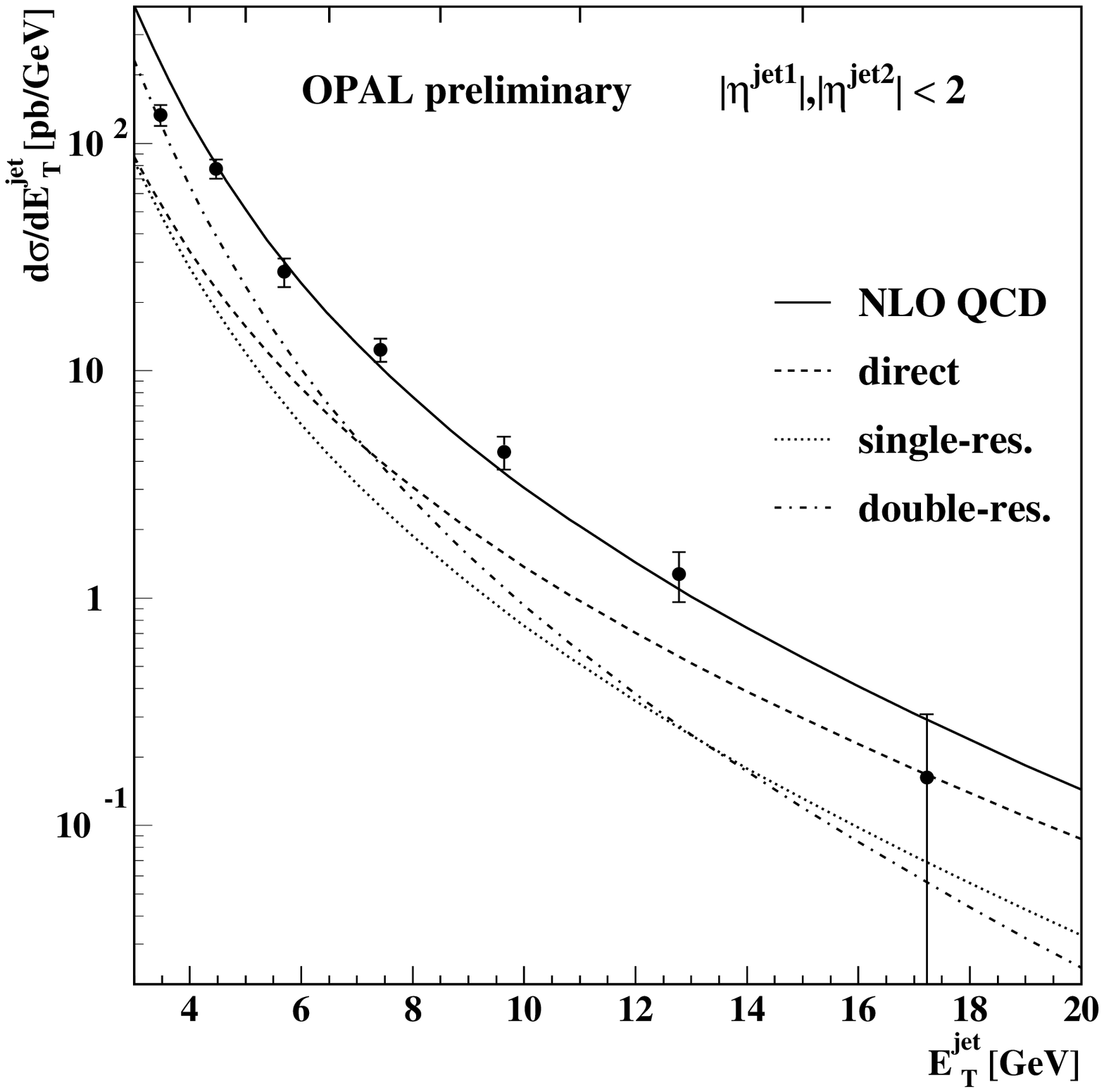,width=0.46\textwidth,height=6.0cm}
 &
\epsfig{file=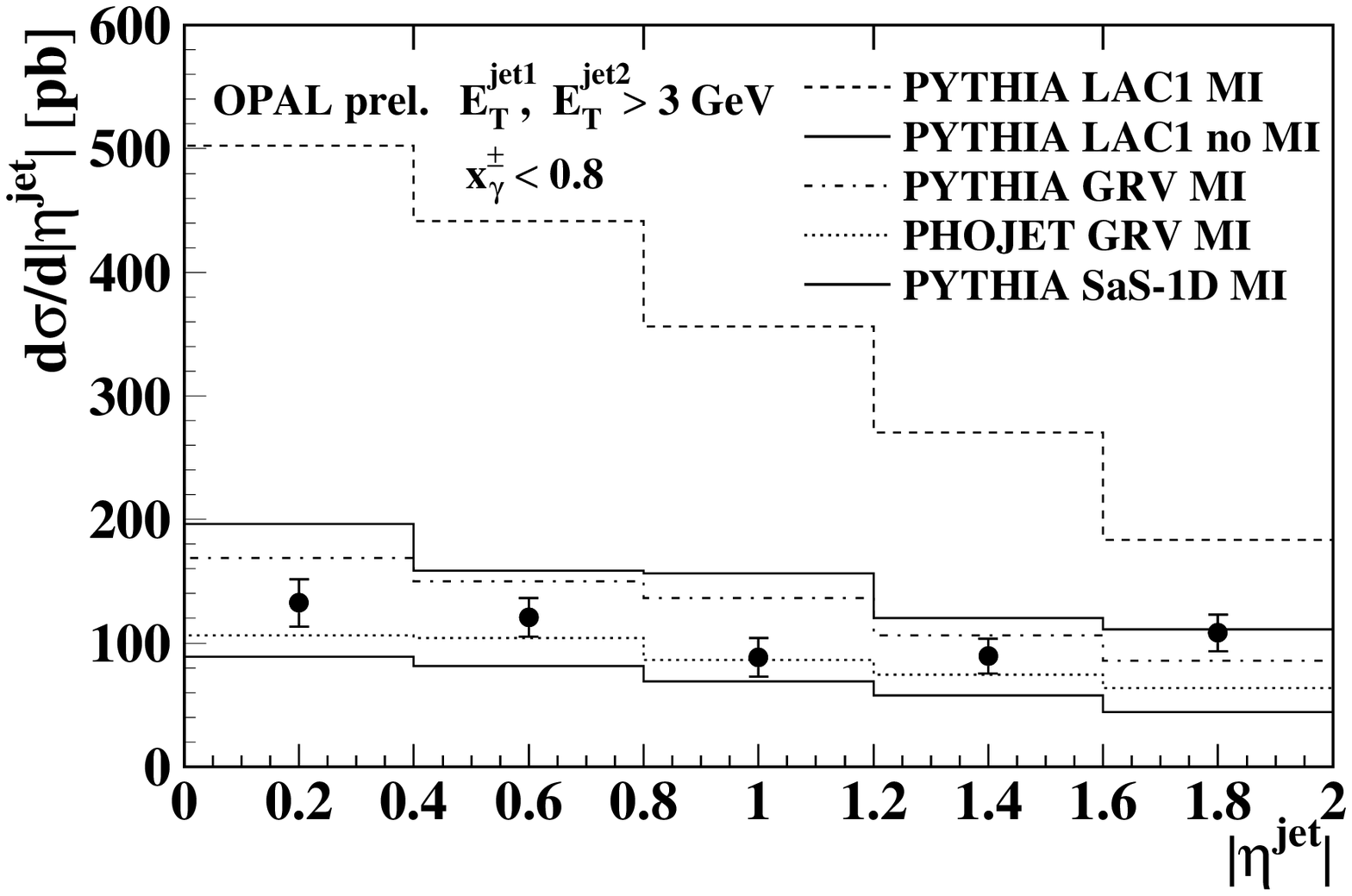,width=0.46\textwidth,height=6.0cm} 
\end{tabular}}
\put(1.5,3.4){(a)}
\put(9.0,3.4){(b)}
\end{picture}
\caption{\label{fig-ettwojet}
a) The inclusive $\ee$ dijet cross-section as a function
of $\ETJET$ for jets with $|\etajet|<2$ using a cone size $R=1$.
b) The inclusive dijet cross-section as a function
of $|\etajet|$ for jets in mainly double-resolved events
with $\ETJET>3$~GeV using a cone size $R=1$.}
\end{figure}

Inclusive one-jet and dijet cross-sections have been
measured in $\gg$ scattering at an $\ee$ centre-of-mass energy of 
$\sqee=58$ GeV at TRISTAN~\cite{bib-amy,bib-topaz} and at $\sqee=130-172$ GeV
by OPAL~\cite{bib-opalgg,bib-opalgg2}. 
The $\ETJET$ distribution for dijet events in the range $|\etajet|<2$ 
measured by OPAL~\cite{bib-opalgg2} at $\sqee=161-172$~GeV
is shown in Fig.~\ref{fig-ettwojet}a. The measurements are
compared to a NLO calculation~\cite{bib-kleinwort} 
which uses the NLO GRV parametrisation~\cite{bib-grv}.
The direct, single- and double-resolved parts and their sum are
shown separately. The data points are in good agreement with
the calculation except in the first bin where
theoretical and experimental uncertainties are large.

To study the sensitivity to the choice of parametrisation
for the parton distributions of the photon, OPAL has also measured
the inclusive dijet cross-section as a function of $|\etajet|$
for events with a large double-resolved 
contribution obtained by requiring $\xgpm<0.8$
(Fig.~\ref{fig-ettwojet}b).
The inclusive dijet cross-section predicted by the two
LO QCD models PYTHIA~\cite{bib-pythia} and 
PHOJET~\cite{bib-phojet} differ significantly even if the same 
parton distribution functions (here LO GRV) are used, 
reducing the sensitivity to the parametrisation.
Different parametrisations were used as input to PYTHIA. 
LO GRV~\cite{bib-grv} and SaS-1D~\cite{bib-sas} describe
the data equally well, but LAC1~\cite{bib-LAC1}, which increases
the cross-section for gluon-initiated processes,
overestimates the inclusive dijet cross-section
significantly. A correct treatment
of multiple parton interactions (MI) is also important. 
PYTHIA plus LAC1 with and without MI 
differs by more than a factor of two. The influence of
MI can be constrained by studying energy flows outside
the jets.

\section{Inclusive charged hadron production}
Measurements of inclusive charged hadron production 
complement similar studies of jet production.
OPAL has measured the differential cross-sections $\dspt$
as a function of the transverse momentum $\pt$
of charged hadrons at $\sqee=161-172$~GeV.
Until now, $\pt$ distributions of charged hadrons have only been
measured for single-tagged events by TASSO~\cite{bib-tasso} and
MARK~II~\cite{bib-mark2} at an average $\langle Q^2 \rangle$
of 0.35~GeV$^2$ and 0.5~GeV$^2$, respectively.

\begin{figure}[htbp]
\begin{picture}(15.0,6.0)
\put(0,3.0){\begin{tabular}{cc}
\epsfig{file=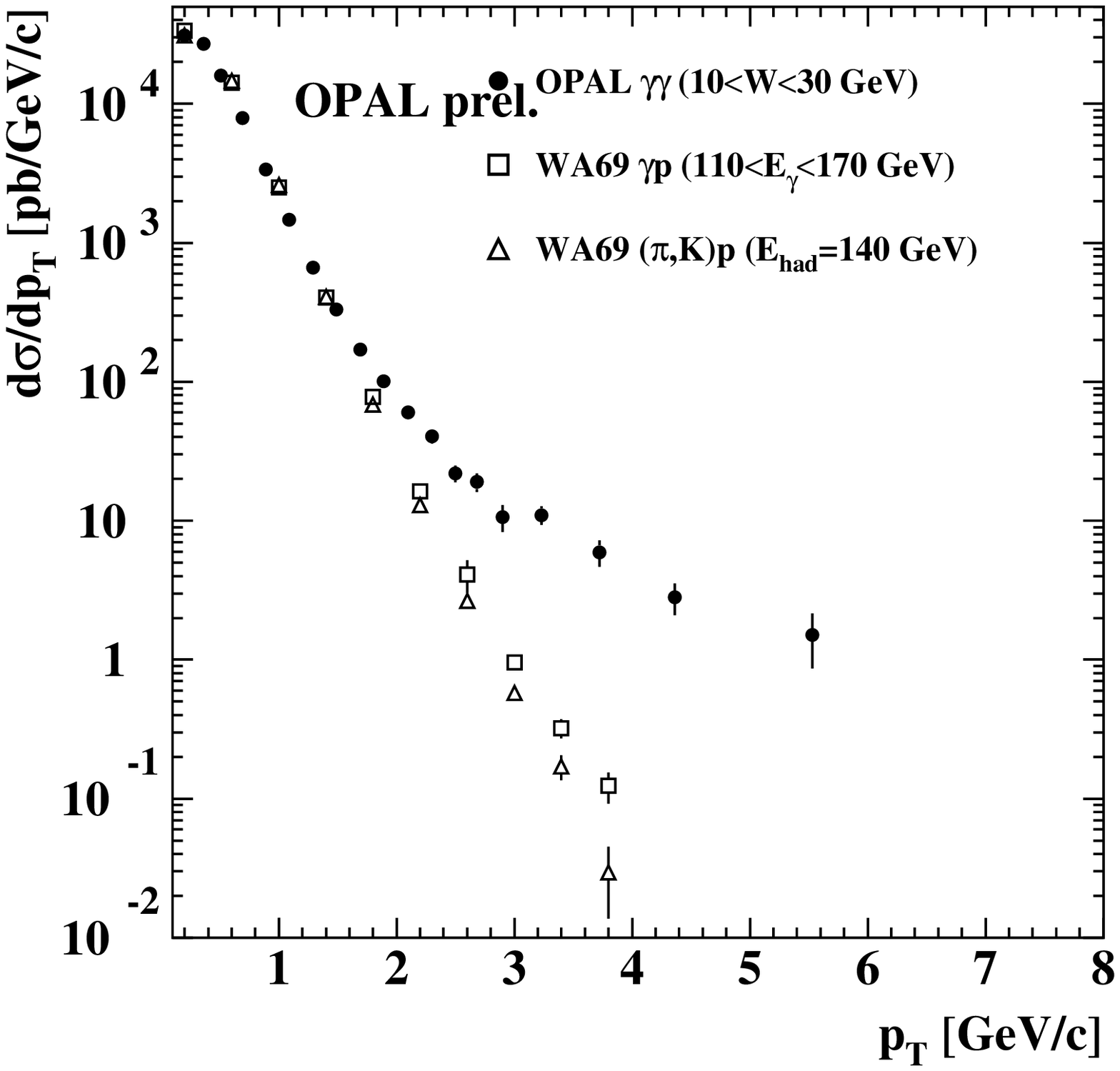,width=0.48\textwidth,height=6.0cm}
 &
\epsfig{file=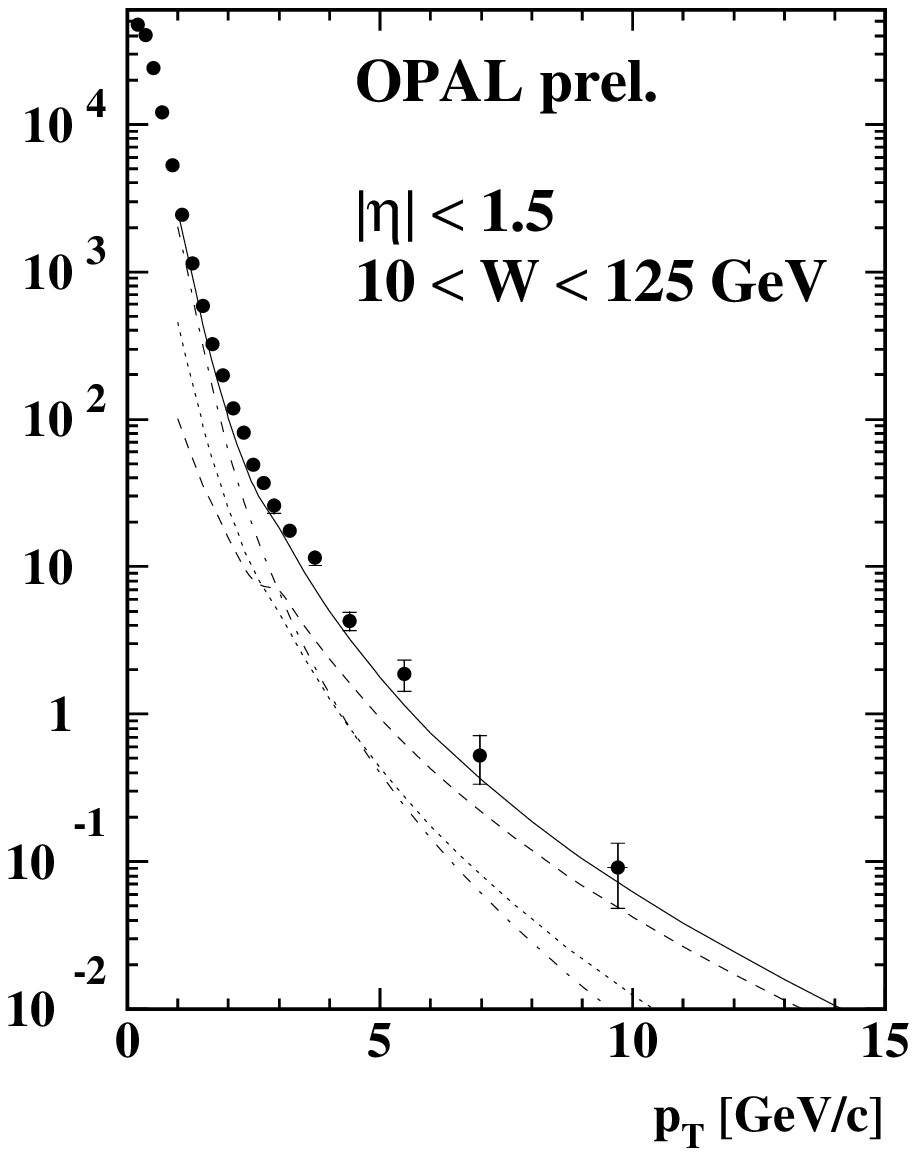,width=0.44\textwidth,height=6.0cm} 
\end{tabular}}
\put(1.5,3.4){(a)}
\put(9.0,3.4){(b)}
\end{picture}
\caption{\label{fig-wa69}
a) The $\pt$ distribution measured in $\gg$ interactions
in the range $10<W<30$~GeV is compared to the $p_{\rm T}$ distribution
measured in $\gamma$p and hp (h$=\pi,$K) interactions by 
WA69~\protect\cite{bib-wa69}. 
The cross-section values given on the ordinate
are only valid for the OPAL data.
b) $\dspt$ for $|\eta|<1.5$ and 
$10<W<125$~GeV compared to NLO calculations for $\pt>1$~GeV$/c$
by Binnewies et al.} 
\end{figure}

The $\pt$ distributions in $\gg$ interactions are expected to
be harder than in $\gamma p$ or hadron-p
interactions due to the direct component. This is
demonstrated in Fig.~\ref{fig-wa69}a by comparing 
the $p_{\rm T}$ distribution for $\gg$ interactions 
to $p_{\rm T}$ distributions
measured in $\gamma$p and hp (h$=\pi,$K) interactions by WA69~\cite{bib-wa69}. 
The WA69 data is normalised to the $\gg$ data in the low 
$p_{\rm T}$ region 
at $\pt\approx 200$~MeV/$c$ using the same factor for the hp and the
$\gamma$p data.
The $p_{\rm T}$ distribution of WA69 has been measured
in the Feynman-$x$ range $0.0<x_{\rm F}<1.0$. The hadronic invariant
mass of the hp and $\gamma$p data is about $W=16$~GeV which is
of similar size as the the average $\langle W \rangle$ of the $\gg$
data in the range $10<W<30$~GeV.
Whereas only a small increase is observed
in the $\gamma$p data compared to the h$\pi$ data at large $\pt$,
there is a significant increase of the relative rate in the range 
$\pt>2$~GeV/$c$ for $\gg$ interactions due to the
direct process. 

The cross-sections $\dspt$ are also compared to NLO
calculations~\cite{bib-binnewies} which
are calculated using the QCD partonic cross-sections
to NLO for direct, single- and double-resolved processes.
The hadronic cross-section is a convolution of the Weizs\"acker-Williams 
effective photon distribution, the parton distribution functions and
the fragmentation functions of Ref.~\cite{bib-bkk} which are obtained
from a fit to $\ee$ data from TPC and ALEPH.
The NLO GRV parametrisation is
used~\cite{bib-grv} with $\Lambda^{(5)}_{\overline{\rm MS}}=131$ MeV
and $m_{\rm c}=1.5$~GeV$/c^2$. 
The renormalization and factorization scales
are set equal to $\xi\pt$ with $\xi=1$. 
The change in slope around $\pt=3$~GeV/$c$ in the
NLO calculation is due to the charm threshold, below which
the charm distribution in the resolved photon and the charm
fragmentation functions are set to zero.

A minimum $p_{\rm T}$ of 1~GeV/$c$ is required to ensure
the validity of the perturbative QCD calculation.
The NLO calculation is shown separately for double-resolved, 
single-resolved and direct 
interactions. At large $\pt$ the direct interactions dominate.
It should be noted that these classifications are scale dependent in NLO. 
The scale dependence of the NLO calculation was studied
by setting $\xi=0.5$ and 2. This leads to a variation of the cross-section
of about $30\%$ at $\pt=1$~GeV/$c$ and about $10\%$ for $\pt>5$~GeV/$c$.
The NLO calculations of $\dspt$ lie about $25\%$ below the data
for $10<W<125$~GeV. 

\section{Total cross sections}
\label{sec-total}
The total cross-sections $\sigma$ for hadron-hadron and $\gamma$p 
collisions are well described by the  parametrisation 
$\sigma=X s^{\epsilon}+Y s^{-\eta}$,
where $\sqrt{s}$ is the centre-of-mass energy of the interaction. 
Assuming factorisation for the 
Pomeron term $X$, the total hadronic $\gg$ cross-section $\sigma_{\gg}$
can be related to the pp (or $\ppbar$) and $\gamma$p total cross-sections at 
high $W=\sqrt{s}_{\gg}$, where the Pomeron trajectory
should dominate:
\begin{equation}
\sigma_{\gg}
=\frac{\sigma_{\gamma{\rm p}}^2}{\sigma_{\rm pp }}.
\label{eq-tot2}
\end{equation}
A slow rise of the total cross-sections with energy is predicted,
corresponding to $\epsilon\approx0.08$.

Before LEP $\sigma_{\gg}(W)$ has been measured
by PLUTO~\cite{bib-pluto}, TPC/2$\gamma$~\cite{bib-tpc} and
MD1~\cite{bib-md1} in the region $W<10$~GeV, before the onset 
of the high energy rise of $\sigma_{\gg}$.
Using LEP data taken at $\sqee=130-161$~GeV
L3~\cite{bib-l3tot} has demonstrated that
$\sigma_{\gg}(W)$ is consistent with the universal Regge behaviour of
total cross-sections in the range $5\le W \le 75$~GeV. The
L3 measurement is shown
\begin{wrapfigure}[19]{r}{0.45\textwidth}
\epsfig{file=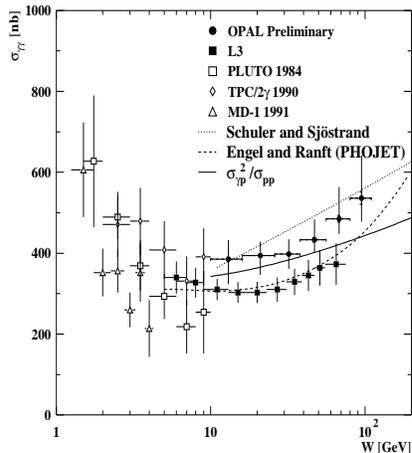,width=0.45\textwidth,height=6.0cm}
\caption{\label{fig-stot} 
Total cross-section of the process $\gg\rightarrow\mbox{hadrons}$
as a function of $W=W_{\gg}=\sqrt{s}_{\gg}$.}
\end{wrapfigure}
in Fig.~\ref{fig-stot}
together with an OPAL measurement in the range
$10<W<110$~GeV using data taken at $\sqee=161-172$~GeV.
The observed energy dependence of the cross-section is similar,
but the values for $\sigmagg$ are about $20\%$ higher. 
The errors are strongly correlated between the $W$ bins
in both experiments. About $5~\%$ discrepancy is due
to the different MC generators used for unfolding.
The origin of the remaining discrepancy is
not yet understood. 

Based on the DL-model~\cite{bib-DL}, 
the assumption of a universal high energy behaviour of 
$\gg$, $\gamma$p  and pp cross-sections is tested.
The parameters $X$ and $Y$ are fitted to the
total $\gg$, $\gamma$p and pp cross-sections 
in order to predict $\sigmagg$ via Eq.~\ref{eq-tot2}
using $\sqrt{s}_{\gg}=\sqrt{s}_{\rm \gamma p}=\sqrt{s}_{\rm pp }$.
The process dependent fit values for $X$ and $Y$ 
and the universal parameters $\epsilon = 0.0790 \pm 0.0011$ and 
$\eta = 0.4678 \pm 0.0059$ are
taken from Ref.~\cite{bib-pdg}.
This simple ansatz gives a reasonable
description of the total $\gg$ cross-section $\sigmagg$. 
The models of Schuler and Sj\"ostrand~\cite{bib-GSTSZP73} and
the model of Engel and Ranft~\cite{bib-phojet} are also shown. 

\section{Conclusions}
In general, $\gg$ interactions are similar to hadron-hadron interactions.
At centre-of-mass energies $\sqrt{s}_{\gg}>10$~GeV the energy
dependence of the total $\gg$ cross-section is consistent 
with the rise observed in hadronic interactions.

QCD is tested in $\gg$ interactions at LEP
by comparing to LO and NLO QCD calculations. 
In case of the photon, the perturbative splitting $\gamma\to\qqbar$
must also be taken into account which modifies the QCD predictions. 
This is observed in the scaling violations of the
photon structure function $F_2^{\gamma}$.

Information about the gluon content of the photon can be extracted
from measurements of jet production. 
NLO calculations are in reasonable agreement with the data.
Comparing transverse momentum distribution
of $\gg$ interactions with hadron-proton or $\gamma$-proton
data shows the relative increase of hard interactions
in $\gg$ processes due to the direct component.

\end{document}